\documentclass[aps,prl,twocolumn,superscriptaddress,showpacs]{revtex4}

\usepackage{graphicx}

\begin{document}

\title{All-Optical Formation of Quantum Degenerate Mixtures}

\author{Takeshi Fukuhara}
\affiliation{Department of Physics, Graduate School of Science, Kyoto University, Kyoto 606-8502, Japan}

\author{Seiji Sugawa}
\affiliation{Department of Physics, Graduate School of Science, Kyoto University, Kyoto 606-8502, Japan}

\author{Yosuke Takasu}
\affiliation{Department of Physics, Graduate School of Science, Kyoto University, Kyoto 606-8502, Japan}

\author{Yoshiro Takahashi}
\affiliation{Department of Physics, Graduate School of Science, Kyoto University, Kyoto 606-8502, Japan}
\affiliation{CREST, Japan Science and Technology Agency, Kawaguchi, Saitama 332-0012, Japan}

\date{\today}

\begin{abstract}
We report the realization of quantum degenerate mixed gases of ytterbium (Yb) isotopes using all-optical methods. We have succeeded in cooling attractively interacting $^{176}$Yb atoms via sympathetic cooling down to below the Bose-Einstein transition temperature, coexisting with a stable condensate of $^{174}$Yb atoms with a repulsive interaction. We have observed a rapid atom loss in $^{176}$Yb atoms after cooling down below the transition temperature, which indicates the collapse of a $^{176}$Yb condensate. The sympathetic cooling technique has been applied to cool a $^{173}$Yb-$^{174}$Yb Fermi-Bose mixture to the quantum degenerate regime. 
\end{abstract}


\maketitle
Ultracold atomic gases have provided deep insight into quantum many-body systems, since the realization of a quantum degenerate gas, such as a Bose-Einstein condensate (BEC) \cite{Anderson95} and a degenerate Fermi gas \cite{DeMarco99}. One of the intriguing new developments in this field is a study of quantum degeneracy in a mixed gas. An advantage in the study using ultracold atomic mixed gases is the ability to select the statistics of atomic gases: a Bose-Bose, Fermi-Bose and Fermi-Fermi mixture. The interactions, which determine the stability and the dynamics of the mixed-gas system, can be also tuned by changing a combination of a mixed gas or by using magnetic and optical Feshbach resonances \cite{Inouye98, Theis04, Enomoto08}. So far, Fermi-Bose mixtures in the degenerate regime have been realized with various combinations of atoms: $^{6}$Li-$^{7}$Li \cite{Truscott99}, $^{6}$Li-$^{23}$Na \cite{Hadzibabic02}, $^{40}$K-$^{87}$Rb \cite{Roati02}, $^{6}$Li-$^{87}$Rb \cite{Silber05}, $^{3}$He-$^{4}$He \cite{McNamara06}. A two-species condensate of $^{41}$K and $^{87}$Rb was created \cite{Modugno02}, and more recently a BEC mixture of Rb isotopes has been reported \cite{Papp08}. 

In order to realize a quantum degenerate mixture of different species, sympathetic cooling in a magnetic trap has been employed, whereas creation of a quantum degenerate mixture in an optical trap has several important advantages. First, this method can be applied to atoms with no permanent magnetic moment, e.g. ytterbium (Yb) and alkaline-earth metal, or atoms unstable in a magnetic trap due to large inelastic collision rates, e.g. Cr \cite{Griesmaier05}. Second, a high density in the tight confinement of the optical trap results in fast production of a quantum degenerate gas. Third, tight confinement of optical traps can reduce a gravitational sag, which keeps the overlap of atom clouds with different masses. So far, sympathetic cooling with two species in optical dipole traps was performed for Li and Cs atoms, but it did not reach quantum degeneracy \cite{Mudrich02}. 

Studies of a degenerate mixed gas using Yb atoms, which have no electron spin in the ground state, are attractive because it has seven stable isotopes: five bosonic isotopes with nuclear spin $I=0$ ($^{168,170,172,174,176}$Yb) and two fermionic isotopes, $^{171}$Yb with $I=1/2$ and $^{173}$Yb with $I=5/2$. Following the realization of a BEC in $^{174}$Yb \cite{Takasu03}, a degenerate Fermi gas of $^{173}$Yb \cite{Fukuhara07a} and a BEC in $^{170}$Yb \cite{Fukuhara07b} has been achieved by evaporative cooling in an optical trap. The s-wave scattering lengths for all combinations of the isotopes have been deduced, and their values distribute over a large range \cite{Kitagawa08}. Moreover, control of the scattering length by using an optical Feshbach resonance has been demonstrated for thermal sample of Yb isotopes \cite{Enomoto08}. The mass of the isotopes are almost the same, which ensures the most efficient thermalization per collision \cite{Mudrich02} and results in a small difference in gravitational sags. Another advantage of Yb is the existence of ultranarrow transitions of $^1$S$_0-^3$P$_0$ and $^1$S$_0-^3$P$_2$. By using this ultranarrow transition, high-resolution spectroscopy to detect small energy difference due to the interspecies interaction can be expected \cite{Yamaguchi08}.

In this paper, we report all-optical creation of quantum degenerate mixtures of Yb isotopes: a Bose-Bose mixture of $^{174}$Yb-$^{176}$Yb and a  Fermi-Bose mixture of $^{173}$Yb-$^{174}$Yb. $2 \times 10^4$ atoms of $^{176}$Yb are cooled down to $\sim 120$nK, which corresponds to the Bose-Einstein transition temperature, coexisting with a BEC of $^{174}$Yb. Further cooling is performed to increase the number of condensate atoms in $^{176}$Yb, but we can not observe growth of the condensate. This fact is consistent with a negative sign of the scattering length of $^{176}$Yb, which has recently been determined from experimental data of high-resolution two-color photoassociation spectroscopy \cite{Kitagawa08}. A rapid atom loss in $^{176}$Yb atoms is observed after evaporative cooling into the degenerate regime, which is attributed to the collapse of the condensate in $^{176}$Yb. The sympathetic cooling technique is also important in terms of cooling fermionic atoms into the quantum degenerate regime. We cool unpolarized and polarized $^{173}$Yb atoms down to 0.7 and 0.4 of the Fermi temperature, respectively, in the presence of a condensate of $^{174}$Yb.

The experiment begins with a simultaneous trapping of two isotopes in a magneto-optical trap (MOT) \cite{Honda02}. Atoms are first decelerated by a Zeeman slower with the $^1$S$_0-^1$P$_1$ transition (the natural linewidth is 28 MHz) and then loaded into the MOT with the $^1$S$_0-^3$P$_1$ transition (the natural linewidth is 182 kHz). Bichromatic MOT beams for simultaneous trapping of two isotopes in the MOT are generated by an electro-optic modulator driven at the frequency corresponding to the isotope shift: 2386 MHz ($^{173}$Yb-$^{174}$Yb) and 955 MHz ($^{174}$Yb-$^{176}$Yb). Each isotope is independently loaded into the MOT; first the frequency of the Zeeman slower laser is tuned to the near-resonance of one isotope, and then is changed to the other isotope. The lifetime of the gas trapped in the MOT is longer than a typical MOT loading time of a few tens of seconds, and thus the atom loss of the first isotope during the loading time for the second isotope is negligibly small. 

After the MOT loading, the laser-cooled mixed gas is loaded into a crossed far-off-resonance trap (FORT) with horizontal and vertical beams, which produces the same conservative potential for isotopes because the isotope shift is negligibly small compared with the detuning of the FORT lasers. The trapping frequencies are also the same within 1\% for isotopes due to the small difference in mass. Particle numbers and temperatures of each gas are obtained using absorption imaging after a ballistic expansion. Isotope-selective absorption imaging is possible because the isotope shift of the $^1$S$_0-^1$P$_1$ transition for probing is much larger than the linewidth. 

Evaporative and sympathetic cooling is performed by continuously decreasing the trap depth of the horizontal FORT beam, while that of the vertical FORT beam is kept constant. Differently from sympathetic cooling in the magnetic trap \cite{Myatt97}, where coolant atoms are selectively cooled by rf or microwave evaporation and target atoms are sympathetically cooled through thermal contact with the coolant atoms, target atoms also evaporate in our case of sympathetic cooling in the optical trap. To realize quantum degenerate gases of target atoms, a sufficiently large number of target atoms should be prepared at the start of evaporation, taking into account decrease in target atoms during evaporation. On the other hand, the larger the number of target atoms, the more coolant atoms evaporate. For this reason, the number ratio of coolant and target atoms at the start of evaporation is important to reach the degenerate regime for the two species, and its optimal rate depends on a combination of the intra- and interspecies scattering lengths. We optimize the ratio of populations by changing the order of the MOT loading and the loading times of two isotopes. 

In order to cool a Bose-Bose mixture of $^{174}$Yb-$^{176}$Yb into the quantum degenerate regime, we prepare $2 \times10^6$ atoms of $^{174}$Yb and $2 \times10^5$ atoms of $^{176}$Yb in the FORT with the MOT loading of $^{176}$Yb for 4 s followed by the $^{174}$Yb loading for 30 s. The evaporation ramp for efficient sympathetic cooling depends on the interspecies collision rate, and we experimentally optimize the ramp. The whole evaporation time optimized for producing the degenerate mixture is 3 s, which is almost the same as that for $^{174}$Yb alone. During the evaporation, the temperature of the $^{176}$Yb cloud is equal to that of the $^{174}$Yb cloud  within the uncertainty of several percent. This indicates that the interspecies collision rate is comparable to or larger than the $^{174}$Yb intraspecies collision rate, which is in agreement with the scattering lengths $a_{174}= +5.55$ nm and $a_{174 \mathchar`- 176}= +2.88$ nm determined by photoassociation spectroscopy \cite{Kitagawa08}. 

\begin{figure}
\begin{center}
\includegraphics[width=0.7\linewidth]{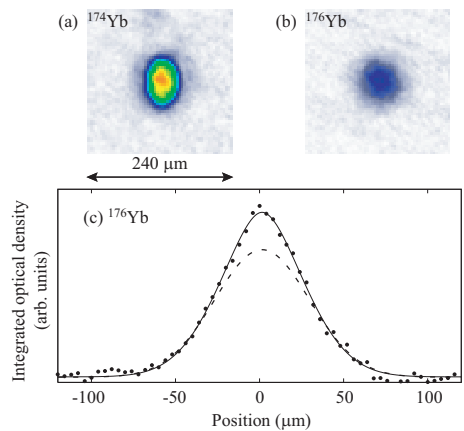}
\end{center}
\caption{(color online). Absorption images of (a) $^{174}$Yb and (b) $^{176}$Yb atoms and density distribution integrated over the vertical direction of $^{176}$Yb atoms after 12 ms of a free expansion. The data are averaged over five measurements. The solid line shows the fit by the Bose-Einstein distribution function and the dashed line shows the Gaussian fit to the wings, yielding a temperature of $T \sim$120 nK.\label{176bimodal}}
\end{figure}
Although the trapping frequencies ($\omega_1$, $\omega_2$, $\omega_3$) are almost the same for the isotopes, the BEC transition temperature $T_c \simeq 0.94 \hbar/k_B (\omega_1 \omega_2 \omega_3 N)^{1/3}$ is different because the number of $^{174}$Yb atoms is larger than that of $^{176}$Yb atoms. Therefore, first the growth of a $^{174}$Yb condensate is observed and then the $^{176}$Yb cloud is cooled down to the transition temperature. Finally, we have observed an almost pure BEC with $6 \times 10^4$ $^{174}$Yb atoms and a non-gaussian momentum distribution of gas with $2 \times10^4$ $^{176}$Yb atoms at 120 nK (Fig. \ref{176bimodal}), where the calculated transition temperature for $^{176}$Yb is 160 nK. The trapping frequencies after evaporative cooling are measured to be $2 \pi \times$(45, 200, 300) Hz by the parametric heating method and by exciting the center-of-mass motion of the $^{174}$Yb condensate alone. At the final stage of evaporation, the trapping potential is strongly affected by the gravity. However, the different gravitational sag for $^{174}$Yb and $^{176}$Yb is calculated to be less than 50 nm. This difference is much smaller than the cloud size of thermal samples and the Thomas-Fermi radius of the $^{174}$Yb condensate along the vertical direction, $\sim$2 $\mu$m. 

Further cooling is performed in order to observe the large number of a condensate in $^{176}$Yb, but the growth of the condensate has not been observed. This result is consistent with a negative scattering length of $^{176}$Yb, $a_{176}= -1.28$ nm \cite{Kitagawa08}, because a condensate with a negative scattering length $- |a|$ becomes unstable to collapse when the number of atoms in the condensate exceeds a critical value $N_{cr} \sim 0.5 a_{ho}/|a|$, where $a_{ho} = \sqrt{\hbar/(m \bar{\omega}})$ is the harmonic oscillator length with the geometrical average $\bar{\omega}=(\omega_1 \omega_2 \omega_3)^{1/3}$ of the trap frequencies \cite{Gerton00, Donley01}. For our trap parameters and the scattering length $a_{176}= -1.28$ nm, the critical value $N_{cr}$ is estimated to be $\sim 250$. The critical number can be modified by the existence of a large stable condensate in another species. Theoretical analysis for the $^{174}$Yb-$^{176}$Yb mixture predicts a decrease in the critical number due to the additional confinement produced by the repulsive interaction with the large stable condensate \cite{Isoshima, Kasamatsu08}. Presently, our imaging system can not distinctly detect such a small number of condensate atoms. By loosening the trap confinement or making the scattering length of $^{176}$Yb smaller via optical Feshbach resonances \cite{Enomoto08}, however, the number of condensate atoms will become larger enough to be clearly observed.

\begin{figure}
\begin{center}
\includegraphics[width=0.6\linewidth]{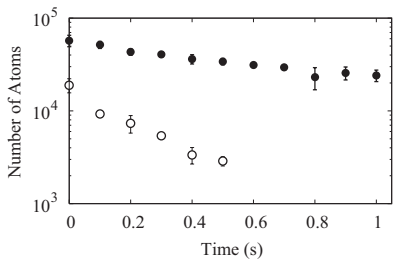}
\end{center}
\caption{Number of $^{174}$Yb (solid circles) and $^{176}$Yb (open circles) atoms after evaporation. Each data point is the average of three measurements. A rapid decrease in the number of $^{176}$Yb atoms indicates that the collapse of a condensate in $^{176}$Yb occurs. \label{lifetime}}
\end{figure}
The stability of the mixed gas in the degenerate regime is investigated by measuring the evolution of the number of $^{174}$Yb and $^{176}$Yb atoms after evaporative cooling (Fig. \ref{lifetime}). First we discuss the decay of $^{174}$Yb atoms. We extract the three-body loss rate constant of $^{174}$Yb from the decay of the $^{174}$Yb condensate alone: $K_{3}^{174}=(4.2 \pm 1.5)\times 10^{-29}$ cm$^6$s$^{-1}$ \cite{Burt97}. We find that the evolution of the number of $^{174}$Yb atoms in the mixture is well described by assuming the $^{174}$Yb three-body loss only. This fact indicates that the three body collisions of $^{174}$Yb-$^{174}$Yb-$^{176}$Yb and $^{174}$Yb-$^{176}$Yb-$^{176}$Yb make negligible contributions to the decay of the mixture. 

On the contrary, the number of $^{176}$Yb atoms decreases by about one order of magnitude in 500 ms. Photon scattering caused by the FORT beams ($\sim 0.2$ Hz) and the loss due to background gas collisions ($< 0.1$ Hz) can be neglected for the short time scale of this decay. Since no binary inelastic collision occurs, the atom loss of $^{176}$Yb is mainly attributed to three-body recombination, such as $^{176}$Yb-$^{176}$Yb-$^{176}$Yb, $^{176}$Yb-$^{176}$Yb-$^{174}$Yb and $^{176}$Yb-$^{174}$Yb-$^{174}$Yb. Since the loss due to the last two collisions should be small as described above, here we focus on the $^{176}$Yb three-body loss only. From a separate measurement on the decay of mixed nondegenerate thermal gases at 400nK, we place a upper limit on the $^{176}$Yb three-body recombination loss rate constant $K_{3}^{176}$ of $3 \times 10^{-28}$ cm$^6$s$^{-1}$. However, the observed loss of the $^{176}$Yb atoms in the degenerate mixture is much larger than atom losses estimated by using the upper bound of the three-body loss rate. Therefore, we consider that the rapid decrease in the number of $^{176}$Yb atoms is due to the growth and collapse of the condensate \cite{Gerton00}. Further investigation on the decay is needed, undoubtedly providing new insights into the stability of degenerate mixed-gases systems. 

The sympathetic cooling technique is also applied for producing a Fermi-Bose degenerate mixture of $^{173}$Yb-$^{174}$Yb. In our previous work, $^{173}$Yb atoms with the spin mixture have been cooled down to quantum degeneracy \cite{Fukuhara07a}. We first conduct simultaneous cooling of a $^{174}$Yb and spin-unpolarized $^{173}$Yb mixture. We collect $9\times10^5$ atoms of $^{173}$Yb and $1 \times10^6$ atoms of $^{174}$Yb in the FORT with the MOT loading of $^{173}$Yb for 30 s and $^{174}$Yb for 20 s. We find no fatal inelastic collisions between the isotopes and therefore $1\times10^4$ $^{173}$Yb atoms and $1 \times10^4$ $^{174}$Yb atoms can be cooled down to approximately 110 nK, which corresponds to 0.7 of the Fermi temperature for $^{173}$Yb and 0.7 of the BEC transition temperature for $^{174}$Yb. The condensate fraction of up to 60\% has been measured. 
\begin{figure}
\begin{center}
\includegraphics[width=0.8\linewidth]{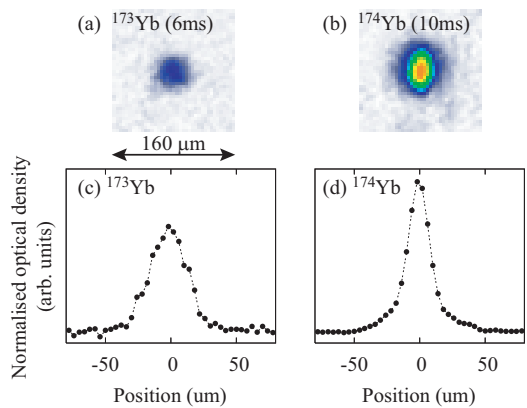}
\end{center}
\caption{(color online). Absorption images ((a) and (b)) and density distributions ((c) and (d)) integrated over the vertical direction of $^{173}$Yb and $^{174}$Yb atoms after a free expansion of 6 ms and 10 ms, respectively. The optical density is normalized by the number of atoms for each sample. The data are averaged over five measurements. \label{173-174}}
\end{figure}

To obtain a further degenerate regime of the $^{173}$Yb Fermi gas, the spin-polarization via optical pumping is effective simply because the population per spin state increases, which results in the higher Fermi temperature. Additionally, the suppression of the inelastic three-body collision involving two or three identical fermions can be expected. After loading into the FORT, a circularly polarized pulse resonant with the $^1$S$_0-^1$P$_1$ ($F_g=5/2 \to F_e=5/2$) transition is applied, to transfer nearly all of the $^{173}$Yb atoms into the $m_{F}=-5/2$ magnetic sublevel. In the pumping process, the atom loss in $^{173}$Yb is less than 20\%. We note that even the polarized $^{173}$Yb atoms alone can be well cooled to a few $\mu$K, which is much less than the p-wave threshold energy of 43 $\mu$K \cite{Fukuhara07a}. This agrees with the existence of a low energy p-wave shape resonance \cite{Kitagawa08}. To achieve the quantum degenerate Fermi-Bose mixture, the polarized $^{173}$Yb atoms are sympathetically cooled through thermal contact with evaporatively cooled $^{174}$Yb atoms. Finally, we produce a degenerate Fermi-Bose mixture of $1 \times10^4$ $^{173}$Yb atoms and $3 \times10^4$ $^{174}$Yb atoms at $\sim$ 130 nK, corresponding to 0.4 of the Fermi temperature (Fig. \ref{173-174}). It is worth noting that by optimizing the cooling scheme to produce deeper Fermi degeneracy of $^{173}$Yb, instead of the Fermi-Bose degenerate mixture, we obtain $4 \times10^4$ $^{173}$Yb atoms at 0.3 of the Fermi temperature.  The Fermi-Bose isotope mixture is an ideal system because almost the same mass suppresses the influence of the differential gravitational sag, which leads to spatial symmetry breaking. Investigations on phase separation \cite{Molmer98} in the Fermi-Bose mixture with a large repulsive interaction \cite{Kitagawa08} and the collective dynamics using such an ideal system can be expected.

We also apply the sympathetic cooling to cool another Yb fermionic isotope $^{171}$Yb. This isotope is a promising sample for fundamental physics, optical lattice clock and quantum information processing because of its ideal structure of spin 1/2. Furthermore, the positive large interspecies scattering length between $^{171}$Yb and $^{174}$Yb \cite{Kitagawa08} is suitable to study spatial separation in a Fermi-Bose mixture \cite{Molmer98}. We successfully cool $^{171}$Yb atoms down to approximately 0.8 of the Fermi temperature, assuming that two spin components are equally distributed. At the final stage of the sympathetic cooling, however, almost all the bosonic $^{174}$Yb atoms evaporate and no further sympathetic cooling can be performed. We consider that improvement of the initial atom number in the optical trap leads to further cooling of $^{171}$Yb into a strongly degenerate regime. 

In conclusion, we have successfully cooled mixed gases of Yb isotopes down to quantum degeneracy via all-optical means. Bosonic $^{176}$Yb atoms is cooled down to below the Bose-Einstein transition temperature, in the presence of a condensate of $^{174}$Yb atoms. A rapid decay of $^{176}$Yb atoms is observed after cooling down below the transition temperature. One possible cause of the decay is the collapse of a $^{176}$Yb condensate. We also cool unpolarized and polarized fermionic $^{173}$Yb atoms down to 0.7 and 0.4 of the Fermi temperature, respectively, coexisting with a BEC of $^{174}$Yb.

We emphasize that mixture experiments using Yb have the advantage that we can easily change Bose-Bose and Fermi-Bose mixtures with the same experimental setup, and also a Fermi-Fermi mixture of $^{171}$Yb-$^{173}$Yb, which is an interesting candidate for the creation of a heteronuclear spinor BCS superfluid \cite{Dickerscheid08}, can be studied. 

We acknowledge T. Isoshima, K. Kasamatsu, and M. Tsubota for helpful discussions, and we thank S. Uetake, K. Enomoto, and A. Yamaguchi for experimental assistance. This work was partially supported by Grant-in-Aid for Scientific Research of JSPS (18204035) and the Global COE Program "The Next Generation of Physics, Spun from Universality and Emergence" from the Ministry of Education, Culture, Sports, Science and Technology (MEXT) of Japan. Y. Takasu acknowledges support from JSPS.

\end{document}